\documentclass[sigconf,natbib=false,authorversion,nonacm=true]{acmart}
\acmConference[CAIN 2024]{3rd International Conference on AI Engineering – Software Engineering for AI}{April 2024}{Lisbon, Portugal}

\copyrightyear{2024}
\acmYear{2024}

\RequirePackage[
  datamodel=acmdatamodel,
  style=acmnumeric,
  ]{biblatex}

\begin{document}

\newcommand{\coo}{\ensuremath{\mathrm{CO_2}}}

\title{Modeling Resilience of Collaborative AI Systems}
\author{Diaeddin Rimawi}
\email{drimawi@unibz.it}
\orcid{0000-0003-3791-399X}
\affiliation{%
  \institution{Free University of Bozen-Bolzano}
  \streetaddress{piazza Università, 1}
  \state{Bolzano}
  \country{Italy}
  \postcode{39100}
}

\author{Antonio Liotta}
\email{antonio.liotta@unibz.it}
\orcid{0000-0002-2773-4421}
\affiliation{%
  \institution{Free University of Bozen-Bolzano}
  \streetaddress{piazza Università, 1}
  \state{Bolzano}
  \country{Italy}
  \postcode{39100}
}

\author{Marco Todescato}
\email{marco.todescato@fraunhofer.it}
\orcid{0000-0003-1449-5692}
\affiliation{%
  \institution{Fraunhofer Italia}
  \state{Bolzano}
  \country{Italy}
  \postcode{39100}
}

\author{Barbara Russo}
\email{barbara.russo@unibz.it}
\orcid{0000-0003-3737-9264}
\affiliation{%
  \institution{Free University of Bozen-Bolzano}
  \streetaddress{piazza Università, 1}
  \state{Bolzano}
  \country{Italy}
  \postcode{39100}
}

\renewcommand{\shortauthors}{Rimawi et al.}

\begin{abstract}

A Collaborative Artificial Intelligence System (CAIS) performs actions in collaboration with the human to achieve a common goal. CAISs can use a trained AI model to control human-system interaction, or they can use human interaction to dynamically learn from humans in an online fashion.
In online learning with human feedback, the AI model evolves by monitoring human interaction through the system sensors in the learning state, and actuates the autonomous components of the CAIS based on the learning in the operational state. 
Therefore, any disruptive event affecting these sensors may affect the AI model's ability to make accurate decisions and degrade the CAIS performance.
Consequently, it is of paramount importance for CAIS managers to be able to automatically track the system performance to understand the resilience of the CAIS upon such disruptive events.
In this paper, we provide a new framework to model CAIS performance when the system experiences a disruptive event.
With our framework, we introduce a model of performance evolution of CAIS. The model is equipped with a set of measures that aim to support CAIS managers in the decision process to achieve the required resilience of the system.
We tested our framework on a real-world case study of a robot collaborating online with the human, when the system is experiencing a disruptive event. 
The case study shows that our framework can be adopted in CAIS and integrated into the online execution of the CAIS activities.
\end{abstract}



\maketitle

\section{Introduction} \label{sec:intro}
\begin{sloppypar}
A Cyber-physical system (CPS) has heterogeneous hardware-software components that collaborate to deliver real-time services. 
The complexity of CPSs is different from one domain to the other.
A Collaborative Artificial Intelligence System (CAIS) is a CPS that performs collaborative actions with humans in a shared environment to achieve a common goal, \cite{bonfanti_gresilience_2023, kadgien_cais-dma_2024}.
Such systems need to be resilient and recover fast from any disruptive events that degrade their performance and eventually prevent them from providing their services within the due schedule, \cite{henry_generic_2012,bonfanti_gresilience_2023}.
\end{sloppypar}

The AI component in CAIS represents the core decision-making instrument that guides system-human interactions. This component can be trained either i) using historical data (offline training), or ii) using data from run-time (online training), \cite{rimawi_green_2022, bonfanti_gresilience_2023,kadgien_cais-dma_2024}.
When the AI component is trained online, the CAIS resilience is of paramount importance, \cite{kadgien_cais-dma_2024}. A disruptive event may indeed affect the ability of the AI component to restore its prediction accuracy in an acceptable time and, in turn, leading to the degradation of overall the system's performance and extra interactions with the human to recover from the event \cite{camilli_microservices_2022}.
Thus, CAISs managers need to ensure their system's resilience by monitoring its performance. Monitoring performance allows the managers to understand if their system was able to detect performance degradation and recover back to an acceptable performance state, \cite{henry_generic_2012, colabianchi_discussing_2021}.

The major goal of this paper is to provide a new framework that models CAIS's performance while learning online and facing a disruptive event. Our model helps in showing the performance evolution of CAIS, the performance degradation that may occur after a disruptive event, and rules and measures to assess CAIS's resilience upon disruptions.
Our framework abstracts the states of CAIS into a \textit{learning state} and an \textit{operational state}. CAIS enters the learning state when first the AI model accuracy is not high enough to make a trusted decision and human interaction is required. Second, when the human intervenes to correct a false positive decision. The CAIS is in the operational state when it completes its service autonomously.
Our framework tracks CAIS's states during run-time and finds the ratio between the number of the operational state over the number of the learning state during a specific time frame, and provide a set of measures to evaluate the performance across the states.
We applied our framework by carrying out an experiment with a real-world CAIS case study.
Our CAIS is a collaborative robot that learns to classify objects based on their colors in an online learning process.
During the experiment, we disrupted the learning process by turning off the supporting lights of the RGB Camera sensor capturing the objects to be classified.
Our framework has shown the incremental evolution of the system performance from a \textit{steady state}, the performance degradation after the disruptive event occurring within a \textit{disruptive state}, and the recovery during a \textit{recovered state}.
The model of resilience we have obtained also shows the system entering into a second disruptive state caused by removing the cause of the disruptive event. After such state again the system enters into recovered state.

Our major contribution in this work can then be summarized as follows:
\begin{enumerate}
    \item We design a novel framework that models the evolution of CAIS performance. Our framework defines the rules and the measurements to automatically track CAIS performance, and detect performance degradation and eventual anomalies.
    \item We define specific measurements and rules to describe the system performance at each state of its evolution. Our rules assess CAIS's resilience upon disruptive events and set the comparison baseline for future research on resilient CAISs.
    \item We automated our framework with a real-world CAIS demonstrator, a collaborative robotic arm with an AI component that online learns from human gestures. We then performed an experiment in a laboratory setting, and we obtained a model of a performance evolution over different states of the system.
\end{enumerate}

The rest of this paper is structured as follows.
In Sec.~\ref{sec:method} we overview the key concepts concerning this work and the proposed methodology. In Sec.~\ref{sec:casestudy} we introduce the research questions, our CAIS demonstrator, our experiment, and the takeaways. In Sec.~\ref{sec:threatstovaility} we discuss threats to validity. In Sec.~\ref{sec:relatedWork} we discuss the related work. Finally, in Sec~\ref{sec:concandfuture} we state our conclusion and the future work.

\section{Methodology} \label{sec:method}
In this paper, we aim to model the resilience\footnote{\textit{Resilience} is a non-functional property that enables a system to recover its performance after an event has degraded it, \cite{henry_generic_2012}.} of a CAIS while it learns online from a human upon disruptive events.
Hence, our method starts by understanding the online learning process, the different variables, and the states a CAIS passes through to achieve resilience. With this knowledge, our method defines a framework and its measures to support CAIS managers in understanding the resilience of their system.
The result is a model as in Fig. \ref{fig:resiliencestates} equipped with measurements as illustrated in the following.

\begin{figure}[th]
\centerline{\includegraphics[width = 0.5 \textwidth]{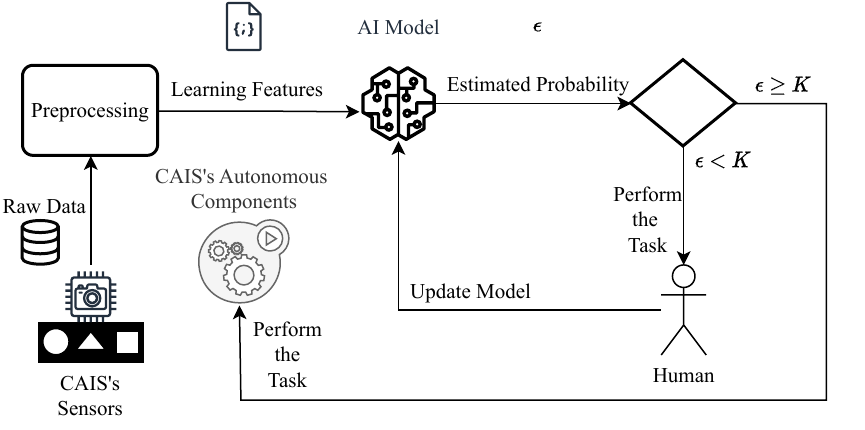}}
\caption{Online Learning Process in CAIS.}
\label{fig:onlinelearningprocess}
\end{figure}

We describe the online learning process by means of the CAIS of our case study.
As such, in the following, we refer to CAIS tasks as specific AI classification tasks, but the online process can be similarly defined for other AI tasks.
Our robot learns object classification by color. Then, based on the learning, the AI component autonomously recommends with a specific probability an action to perform the classification.
Fig.~\ref{fig:onlinelearningprocess} shows the online learning process in a CAIS, where the data is collected by the CAIS sensors, and then preprocessed to extract the learning features. 
With the learning features, the AI model estimates the probabilities of the classification classes and computes their maximum (a.k.a. the confidence level of prediction),  $\epsilon \in [0, 1]$ so that the class with the highest probability is chosen by the robotic arm. To avoid cases in which more than one class can be chosen as the probabilities are similar, the desired confidence level $K$ is set (e.g., $K=0.4$ for three classes). 
The value of $\epsilon$ is then compared with  $K$. If $\epsilon$ is less than $K$, the human is prompted to perform the task, otherwise, the CAIS autonomously does it. Moreover, the human has the possibility to intervene the robot misclassifies an object  (false positives) switching the robot to the ``learning from human" mode.

\subsection{Performance Evolution} \label{subsec:perfevolution}
AI models evolve with learning data. In our context, an AI component starts without any knowledge about the action to perform, which means that $\epsilon$ is zero. 
As CAIS keeps running, the accuracy of the AI model increases with $\epsilon$ in a positive relationship.
With higher values of $\epsilon$, the number of autonomous actions in a time frame increases, and, correspondingly, the number of human actions decreases. The ratio between these two numbers characterizes the interaction between the autonomous components in a CAIS and the human. The goal of CAIS is to keep this ratio at its maximum.
When CAIS experiences a disruptive event, the data and the online learning process might be affected. Therefore, with a disruptive event, the value of $\epsilon$ and the model accuracy may decrease. When the $\epsilon$ value decreases below $K$, the ratio between the number of autonomous actions and the number of human actions decreases, indicating a degradation of the CAIS performance.
Hence, for CAIS's AI model to be able to recover back its ability to perform its tasks autonomously, it requires further training by switching to the learning mode.
The performance curve shows the performance degradation caused by the disruptive event, and recovering back to an acceptable performance level. This evolution shows evidence about CAIS resilience upon the current disruptive event, as described in the following.

\subsection{Modeling Performance} \label{subsec:modelingperformance}
\begin{figure}[th]
\centerline{\includegraphics[width = 0.5 \textwidth]{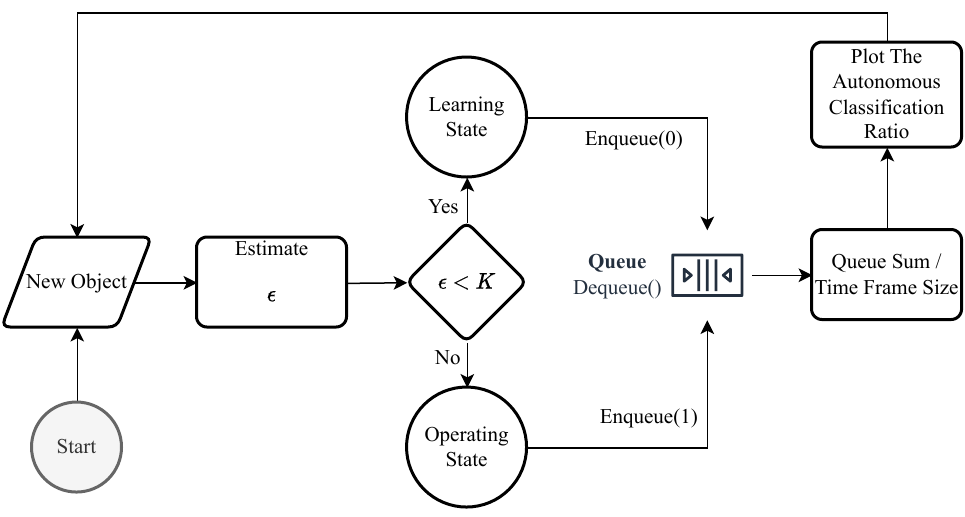}}
\caption{Framework execution states.}
\label{fig:modelingperformance}
\end{figure}

To model the CAIS performance during learning online from the human, we plot the ratio between the number of times CAIS operates autonomously and the number of times the human operates in a specific time frame. Fig.~\ref{fig:modelingperformance} shows the ratio over the states of the system. Our model initiates a first-in first-out queue with a time frame size of zeros. When a new object arrives, CAIS's AI model estimates the classification probability ($\epsilon$), then it enqueues a zero if $\epsilon < K$ (learning state), and one wait to complete the operating state to enqueue one (to allow the human to intervene in case of false positives). After enqueueing a new value, our model dequeues the value on top of the queue to keep only a time frame size elements. Then we define our measurement, the \textit{Autonomous Classification Ratio (ACR)} as the ratio of the autonomous actions in a time frame. We compute the ACR by finding the queue sum over the time frame size. For each new object, we plot the ACR value, which results in the final performance model.

\begin{figure*}[th]
\centerline{\includegraphics[width = 0.7\textwidth]{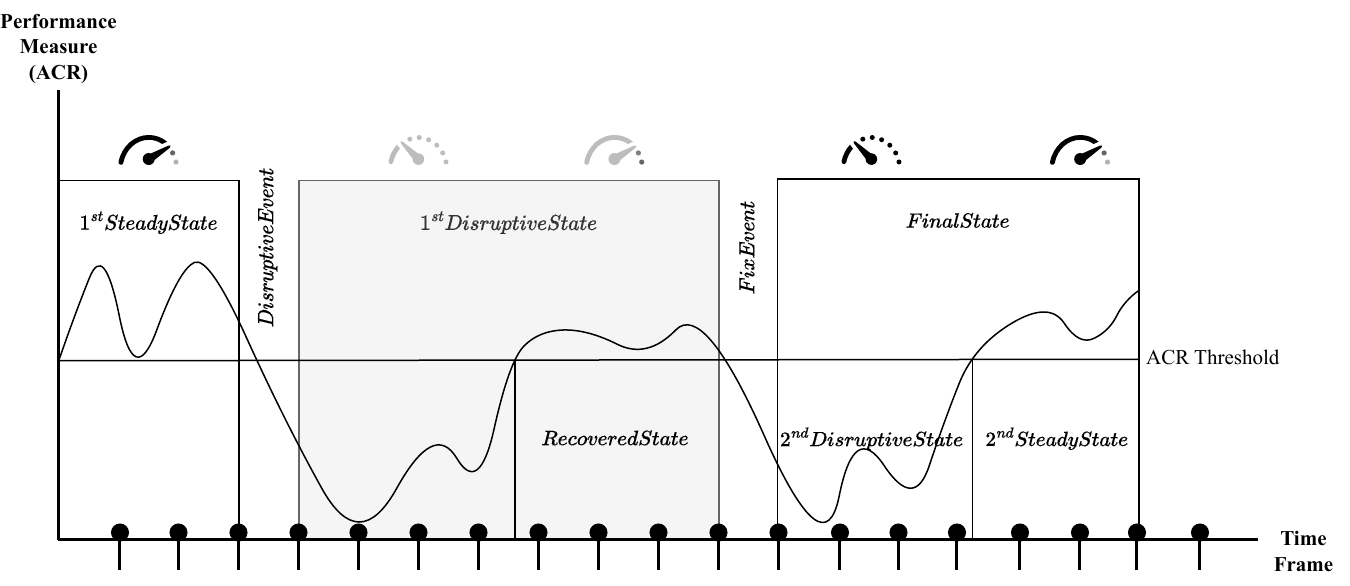}}
\caption{CAIS Resilience Model.}
\label{fig:resiliencestates}
\end{figure*}

Fig.~\ref{fig:resiliencestates}, shows the expected resilience model of CAIS performance evolution when encountering a disruptive event. CAIS learns the classification until it enters the \textit{$1^{st}$ Steady State}. The start of this state is indicated by a set of autonomous classifications for a whole time frame ($ACR = 1$). In the $1^{st}$ Steady State, we define the \textit{ACR Threshold}, which is the minimum ACR value. This value represents the level of performance when the AI component is learning autonomously. 
A value of $ACR = 0$ indicates the end of the $1^{st}$ Steady State and the start of the \textit{$1^{st}$ Disruptive State} after a disruptive event has occurred. During the $1^{st}$ Disruptive State and due to the policies of the online learning process, CAIS aim to learn again to restore the performance to an acceptable level ($ACR \ge ACR\,\,Threshold$). When the level is reached, the CAIS enters into  the \textit{Recovered State}. The recovery at this state, represents the system resilience during disruption.
After recovery, the cause of the disruptive event can be removed and the CAIS enters with the same modalities as above in a \textit{Final State}, which again includes a \textit{$2^{nd}$ Disruptive State} and a \textit{$2^{nd}$ Steady State}. In the Final State, the CAIS may behave differently than in the previous disruptive and steady states, as it may maintain some historical memory of the disruptive event. It is important to note that this cycle of states are repeatable per each disruptive event that degrades the system performance. Table~\ref{tab:statesdef} summarizes each of the performance states of CAIS to represent resilience.

\begin{table}[th]
\caption{Performance States Definitions.}
\begin{center}
\begin{tabular}{p{1in}p{2in}}
\hline
  \textbf{State Name} & \textbf{Definition} \\ \hline
  $1^{st}\,\,Steady\,\,State$ & After CAIS learns enough to start making decisions autonomously with a higher rate than requiring the human. \\ 
  $ACR\,\, Threshold$ & The minimum ACR value during the $1^{st}\,\, Steady \,\,State$. \\ 
  $1^{st}\,\, Disruptive \,\,State$ & Starts after a disruptive event has occurred, leading to performance degradation.  \\ 
  $Recovered \,\,State$ & When a performance degradation is restored back to an acceptable value $ACR > ACR\,\,Threshold$. \\ 
  $Final \,\, State$ & Starts after the removal or fix of the disruptive event source.  \\ 
  $2^{nd}\,\,Disruptive\,\,State$ & The performance degradation happens due to the addiction to the disruptive event. \\ 
  $2^{nd} \,\,Steady \,\,State$ & Recover back CAIS performance from degradation to an acceptable performance level. \\ \hline
\end{tabular}
\label{tab:statesdef}
\end{center}
\end{table}

We have defined a few measures (Table~\ref{tab:rules}) and rules to support CAIS managers in using our framework and assessing the CAIS resilience, as described in the following. The \textit{State Length} is an experimental rule that defines the minimum number of iterations (each new object initiate a new iteration) of a state. We set the state length as equal to the number of iterations in the $1^{st}$ Steady State. Then, in a disruptive state, we examine a period of close to or equal to this number of iterations, starting from the last ACR point under the ACR Threshold, to understand whether CAIS has recovered or not.
Secondly, the \textit{Points Under the Threshold (PUT)} and the \textit{Points Above the Threshold (PAT)}. Each of these measures indicates how well CAIS resilience is. The goal of CAIS's managers is to maximize the PAT ratio over the PUT.
Finally, from a business perspective, a goal of CAIS managers is to minimize the human efforts. Thus, we define the \textit{Human Interaction Ratio (HI Average)} during the disruptive state to reduce human efforts. We defined the HI Average as the average of the human interactions to classify an object during the disruptive state.

\begin{table}[th]
\caption{Resilience Rules and Measurements.}
\begin{center}
\begin{tabular}{p{0.9in}p{2.1in}}
\hline
  \textbf{Criteria} & \textbf{Definition} \\ \hline
  State Length & The length of the $1^{st}$ Steady State, which defines the minimum number of iterations to identify the recovered state. \\ 
  PUT & The number of ACR points under the ACR Threshold in the disruptive state. \\ 
  PAT & The number of ACR points above the ACR Threshold in the disruptive state.  \\ 
  PUT to PAT Ratio & The ratio between the PUT to the PAT.  \\ 
  HI Average & The average of human interactions' in the disruptive state.  \\\hline
\end{tabular}
\label{tab:rules}
\end{center}
\end{table}

\section{Case Study} \label{sec:casestudy}
\begin{figure}[ht]
\centerline{\includegraphics[width = 0.5 \textwidth]{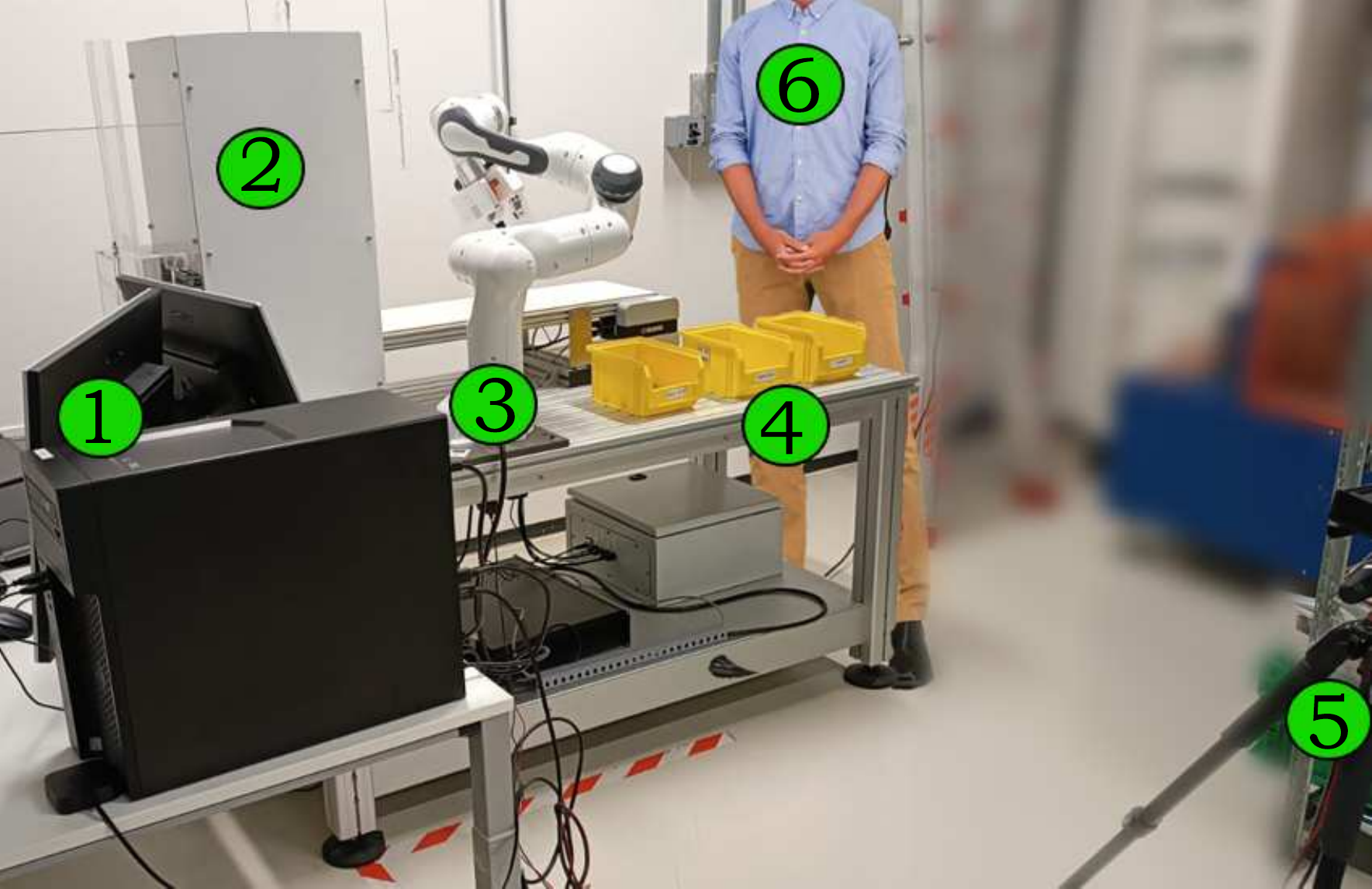}}
\caption{Collaborative Robot Learning from Demonstrations.}
\label{fig:coral}
\end{figure}

In this paper, we introduce our real-world CAIS case study. The CAIS demonstrator, shown in Fig.~\ref{fig:coral}, is a robotic arm (3) responsible for classifying objects based on their color.
The robot learns the object box (4) by tracking the human movement (6) and mapping it with its color histogram.
The robot has two vision sensors, one is placed above the conveyor belt (2), and another (5) tracks the human skeleton. The two vision sensors communicate with computer vision software (1).

To understand if CAIS is resilience upon disruptive events by modeling CAIS performance, we aim to answer the following research questions:
\begin{itemize}
    \item \textbf{RQ1.} \textit{How does CAIS's performance in an online learning process evolve?} To answer this question, we will show a rendered plot of CAIS performance learning from the human, to explain the performance evolution in each state.
    \item \textbf{RQ2.} \textit{What are the rules and measures that indicate if CAIS is resilient upon disruptive events or not?} To answer this question, we will examine the set of measurements and rules to tell if the resulting performance represents a resilient behavior or not. Additionally, these measurements can set the baseline for future research that aims to create more resilient CAISs.
\end{itemize}

To address these questions, we perform an experiment where we automate our framework to track the performance evolution of our CAIS in classifying objects of three colors (Red, Green, and Blue). 
The robot learns the three classes (Box1, Box2, and Box3) from tracking the human hand movements and gestures. 
We enforce a disruptive event to cause a performance degradation to the classification learning process by turning off the supporting lights of the RGB Camera (Fig.~\ref{fig:coral} (2)).
The experiment then runs in iterations, where each iteration represents a new object entered to be classified. 
To avoid the learner being trained for one color more than the other, the objects will enter the conveyor belt sorted red, green then blue, with the same quantity.

\subsection{Experiment Execution} \label{subsec:demo}

The experiment is executed in iterations and for each iteration, we collect 
the AI prediction probability ($\epsilon$), and the iteration state (learning/operating state).
To choose the value of $K$, we consider the three classes we have ($1 / 3 \simeq 0.33$). Thus, we chose a value that is too low ($K = 0.33$), which led to closer values of classification probabilities, and thus, we had two classes with almost the same classification probability. On the other hand, we chose a high-value ($K = 0.50\%$), which required a longer time of training to reach a steady state. The value that best fits our needs is ($K = 0.40$), which helps avoid the two problems.

\begin{figure*}[th]
\centerline{\includegraphics[width = 0.7 \textwidth]{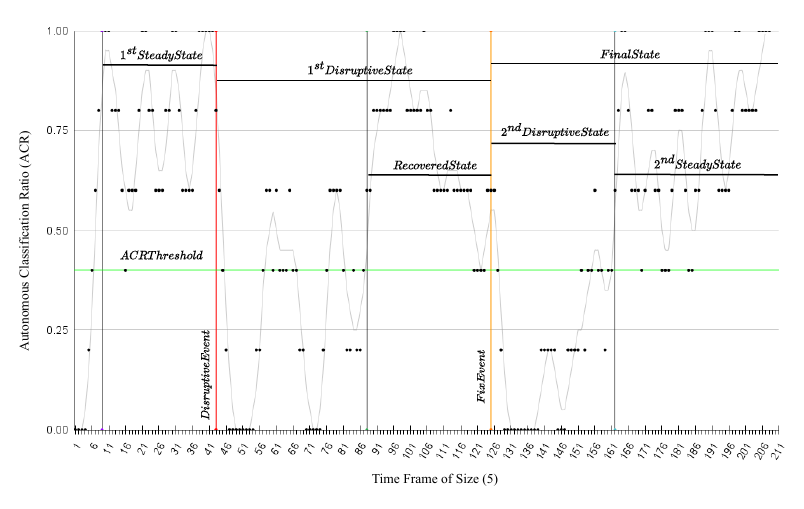}}
\caption{Modeling of CAIS Performance Behavior.}
\label{fig:benchmark}
\end{figure*}

During the execution, CAIS performance is expected to transit through different states in which the AI component online learns its task. 
Fig.~\ref{fig:benchmark} shows the performance evolution of our CAIS demonstrator (collaborative robot) as a function of time frame, where each time frame size is equal to five overlapping iterations (or five objects). The figure shows the ACR values over 208 iterations, which shows all the different performance evolution states.
We triggered a disruptive event by switching off the lights over the conveyor belt supporting the RGB Camera. 
This disrupts the robot's vision and the color histogram extracted from the object image. 
In this experiment, our robot was able to recover both when we switched off and back on the light.
The performance evolution model shows that the system will manage to recover back to a recovered state during the disruptive state. It  will also be able to go back to a steady state after removing the cause of the disruptive event.

\subsection{Takeaways} \label{subsec:takeaways}

In this experiment, we equipped our real-world demonstrator with our framework. The results of the performance evolution model help address our research questions as follows:

\noindent\textbf{Answer to RQ1.} The resilient behavior of CAIS is its ability to overcome unforeseen disruptive events and recover its performance from a degradation state to an acceptable state. Fig.~\ref{fig:benchmark} shows the performance evolution of our CAIS demonstrator (collaborative robot). The performance evolution shows our robot starting with $ACR = 0$, and then starting to learn from the human gestures until entering the $1^{st}$ Steady State. In the $1^{st}$ Steady State, the performance formulates a pattern of going up and down, which is expected due to slight changes of the environment (for example the object position on the conveyor belt). 
The ACR Threshold is calculated to (0.40), which is the minimum value in the $1^{st}$ Steady State.
Then the disruptive event was enforced, and the performance degraded to $ACR = 0$ indicating the beginning of the $1^{st}$ Disruptive State. During the $1^{st}$ Disruptive State, our robot starts learning again until it manages to recover its performance and enter the Recovered State. 
To complete the cycle, we removed the source of the disruptive event (Fix Event - turn back the light on), where we trigger a $2^{nd}$ Disruptive State in the Final State. The robot has replicated the same behavior and learned to recover back into a $2^{nd}$ Steady State.
Our experiment has shown the same performance evolution as the expected performance evolution in Fig.~\ref{fig:resiliencestates}. 

\noindent\textbf{Answer to RQ2.} The answer to this question shows the feasibility of our rules and measures defined in Table~\ref{tab:rules}.
First, the $State\,\,Length$, in our experiment, the length of the $1^{st}$ Steady State was $34$ iterations, thus when the $1^{st}$ Disruptive State started, we had to wait for at least $34$ iterations after the last point under the threshold. This results in a Recovered State of length $37$, and the $2^{nd}$ Steady State, $40+$ since it is the last state.
The experiment shows a total number of $PUT = 23$, and the $PAT = 59$, which makes the ratios equal $0.28$ PUT to $0.72$ PAT, the higher ratio of PAT the better.
Finally, for the $HI Average$ we need to find the number of times CAIS requires a human classification, in other words, the number of times entering the learning state. In our experiment, the number of human interactions was equal to $44$ and the number of iterations is $82$, thus $HI Average = 44 / 82 = 0.54$.
These measures summarize the ability of our CAIS to be resilience to the disruptive event of turning off the lights.

\section{Threats to Validity} \label{sec:threatstovaility}
The major threat to the validity of our study is the internal, conclusion, and external threats. 
\noindent\textbf{Internal Validity.} is related to the factors affecting the research findings. In our experiment, we had to set the size of the time frame to compute ACR. We use a window size that is not too small and not too long, which is five. However, we plan to run a stress experiment with various sizes of time frames to check the effect on the final results.
\noindent\textbf{Conclusion Validity.} concerns the appropriateness of the measurements analysis and inferences drawn from the data. Our resilience rules and measurements are constructed from our observation of our experimental protocol. However, to have a meaningful representation, we consider them to be comparison rules for future research studies to enhance resilience in CAIS.
\noindent\textbf{External Validity.} is related to the degree of support for the generalization of the theoretical results. The nature of our research is mainly exploratory, and generalization beyond our case study is needed for consolidating our claims. Thus, we plan to run more experiments with another real-world demo we have in-house and with simulators.

\section{Related Work} \label{sec:relatedWork}
We have reviewed existing literature according to i) resilience and ii) online learning in the context of CAIS. In the following, we overview them. 

\noindent\textbf{Resilience.}
The state of the art has proposed several models to address resilience, like \textcite{januario_distributed_2019} using a multi-agent model,\textcite{liu_distributionally_2022} using a tri-optimization model, and \textcite{zarandi_detection_2020} using a deep learning model, to mention a few.
All proposed models aim to detect performance degradation or the disruptive event that caused it and return the system to an acceptable performance state.
Disruptive events vary from one system to the other, for example, the disruptive event can be a security vulnerability of the system \cite{zarandi_detection_2020, liu_distributionally_2022}, defect in the system cyber or physical parts \cite{januario_distributed_2019}, from an oracle component of the system such as humans \cite{bonfanti_gresilience_2023}, or it can come from the system surrounding environment, \cite{henry_generic_2012}.

\noindent\textbf{Online learning.} CAIS collaboration with humans is a process of transferring knowledge from the human to the autonomous components of CAIS. \textcite{wang_facilitating_2018} has defined three stages for human-system learning start teaching the robot through voice instructions, the robot then learns, and finally, the human and robot collaborate. However, updating the learning will require us to manually turn to the learning stage, which is not the case in our dynamic continuous online learning.
A recent survey of learning strategy for robot-human collaboration by \textcite{mukherjee_survey_2022} has discussed several input modes, such as gaze, gesture, voice commands, and facial emotions. Specifically, our system uses human gestures for object classification, which requires scene understanding, \cite{mukherjee_survey_2022}. The survey illustrates that the majority of the work is about the safety and physical safety of the human-system collaboration. 

\section{Conclusion and Future Work}
\label{sec:concandfuture}
\textbf{In conclusion}, this work proposes a novel framework to model the performance evolution of CAIS, by tracking the different states of the online learning process with the human. Additionally, it defines the measures and rules to assess CAIS resilience over time. Finally, it automatizes the framework and its measures by designing and executing an experiment with a real-world collaborative robot. The results show the framework's ability to render the performance evolution of our CAIS passing through the different performance states (steady, disrupted, and recovered).

\noindent\textbf{In future work}, we plan to execute additional experiments to generalize our results to other real-world and simulated case studies. We are also designing comparative experiments to automatically support CAIS in decision-making, to enhance CAIS resilience based on the defined measures of this study. Moreover, we plan to reconsider further human attributes, for example human energy.

\begin{acks}
This work is supported by the doctoral fellowship of the National Operational Program (PON) Research and Innovation 2014-2020 (CCI 2014IT16M2OP005), ESF REACT-EU resources.
\end{acks}

\printbibliography

\end{document}